\documentclass[12pt]{article}
\usepackage[cp1251]{inputenc}
\usepackage{amsfonts}
\usepackage[english]{babel}
\usepackage{eufrak}
\usepackage{cite}
\usepackage{amssymb}

\begin{document}
\title{Nonequlibrium Renormalization Theory I}

\author{D.V. Prokhorenko}
\maketitle

\sloppy
\begin{abstract}
In the present article we consider some general class of divergent
diagrams in Keldysh diagram technique. These divergences arise for
non-equilibrium matter and closely related to the divergences in
the kinetic theory of gases. We suggest a scheme of
renormalization of such divergences and illustrate it on some
examples. In the other papers of these series we develop the
general theory of renormalization of non-equilibrium diagram
technique. The fact that thermal divergences in non-equilibrium
diagram technique can be renormalized leads to the following
consequence: to prove that the system tends to the thermal
equilibrium one should to take into account the behavior of the
system on its boundary. In this paper we illustrate this fact on
Bogoliubov derivation of kinetic equations.
\end{abstract}
 \newpage
 \section{Introduction}
There are exist divergences in virial decomposition of kinetic
equations. This fact was observed by Cohen and Dorfman \cite{1}. It is possible to summize
some set of diagrams to obtain a finite result. It was done by Kawasaki and Oppenheim \cite{2}.

Our main goal in this series of papers is to analyze such
divergences. In the present paper we study divergences in Keldysh
diagram technique which arise if the state of the matter is
non-equilibrium. It is more or less obvious that these divergences
are the same as the divergences in the kinetic equations.

In the present series of papers we develop the general theory of
such divergences. As result for a wide class of Bose systems in
the sense of formal power series on coupling constant we find
non-Gibbs state \(\langle \cdot \rangle\) such that the
correlators
\begin{eqnarray}
\langle\Psi^\pm(t,{x}_1)...\Psi^\pm(t,{x}_n)\rangle
\end{eqnarray}
are translation invariant, do not depend on \(t\) and satisfy to
the weak cluster property. Here \(\Psi^\pm\) are the fields
operators and the weak cluster property means the following
\begin{eqnarray}
\lim_{|{a}|\rightarrow\infty}\int
\langle\Psi^\pm(t,{x}_1+\delta_1 e_1{a})
...\Psi^\pm(t,{x}_n+\delta_n e_1 {a} )\rangle f({x}_1,...,{x}_n)d^3x_1...d^3x_n\nonumber\\
=\int\langle\Psi^\pm(t,{x}_{i_1})...\Psi^\pm(t,{x}_{i_k})\rangle
\langle\Psi^\pm(t,{x}_{i_k})...\Psi^\pm(t,{x}_{i_n})\rangle
\nonumber\\
\times f({x}_1,...,{x}_n)d^3x_1...d^3x_n,
\end{eqnarray}
there \(\delta_i\in\{1,0\},\;i=1,2...n\) and
\begin{eqnarray}
i_1<i_2<...<i_k,\nonumber\\
i_{k+1}<i_{k+2}<...<i_n,\nonumber\\
\{i_1,i_2,...,i_k\}=\{i=1,2...n|\delta_i=0\}\neq\emptyset,\nonumber\\
\{i_{k+1},i_{k+2},...,i_n\}=\{i=1,2...n|\delta_i=1\}\neq\emptyset.
\end{eqnarray}
\(f({x}_1,...,{x}_n)\) is a test function, \(e_1\) is a unite vector parallel to the \(x\)-axis.
 In the present paper we consider only a
some wide class of divergent diagrams. 

Let us prove that the existence of such states implies non-ergodic property of the system.
We consider the problem only on classical level. Suppose that our system is ergodic, i.e. there no
first integrals for the system except energy. Then, the distribution function depends only on energy.
We can represent the distribution function \(f(E)\) as follows:
\begin{eqnarray}
f(E)=\sum c_\alpha \delta(E-E_\alpha),
\end{eqnarray}
where the sum can be continuous (integral). Let 1 be some finite subsystem of our system.
Let 2 be a subsystem obtained from 1 by translation on the vector \(L\) parallel to the \(x\)-axis
of enough large length. Let 12 be a union of the subsystems 1 and 2. Let \(\rho_1\), \(\rho_2\) and \(\rho_{12}\)
be distribution functions for the subsystems 1,2 and 12 respectively. Let \(\Gamma_1\), \(\Gamma_2\) and
\(\Gamma_{12}\) be points of the phase spaces for the subsystem 1, 2 and 12 respectively. By the same method
as the method used for the derivation of the Gibbs distribution we find:
\begin{eqnarray}
\rho_{12}=\sum c_\alpha\frac{e^{-\frac{E_{\Gamma_1}}{T}}}{Z_\alpha}\frac{e^{-\frac{E_{\Gamma_2}}{T}}}{Z_\alpha}
\end{eqnarray}
 in the obvious notation. But the weak cluster property implies that
 \begin{eqnarray}
 \rho_{12}=\rho_1\rho_2.
 \end{eqnarray} 
 Therefore all the coefficients \(c_\alpha\) are equal to zero except one. We find that
 \begin{eqnarray}
 f(E)=c\delta(E-E_0)
 \end{eqnarray}
 for some constants \(c\) and \(E_0\). So each finite subsystem of our system can be described by Gibbs
 formula and we obtain a contradiction.
 
Non-ergodic property means that there no thermalization in infinite Bose-gas system.

This fact implies that to prove that the system tends to thermal
equilibrium we should to take into account the behavior of the
system on its boundary. Indeed if the system has no boundary the
system is infinite.

To illustrate this fact we will study Bogoliubov derivation of
kinetic equations. When one derives BBGKI-chain one neglects by
some boundary terms. If one take into account this boundary terms
and use Bogoliubov method of derivation of kinetic equations one
find that these boundary terms compensate the scattering integral.

The paper composed as follows. In section 2 we introduce the
notions of the algebra of canonical commutative relations. In
section 3 we describe our model. In section 4 we describe
non-equilibrium (Keldysh) diagram technique. In section 5 we
discuss the divergences in our model and the method of its
renormalizations. In section 6 we give a proof that there exists
divergences in our theory. In section 7 we describe regularization
which will be used. In section 8 we discuss some simple relation
on Green functions. In section 9 we make renormalization procedure
in one-chain approximation. In section 10 we begin to renormalize
diagrams in two-chain approximation. We calculate divergent parts
of all diagrams in this approximation. In section 11 we discuss
subdivergences i.e. we calculate the contributions which comes
from counterterms for the one-chain diagram. In section 12 we show
that the divergent part of all diagrams which is proportional to
\(\frac{1}{\varepsilon^2}\delta(\omega-\omega(p))\) can be
subtracted by counterterms. In section 13 we show that the
divergent part of all diagram which is proportional to
\(\frac{1}{\varepsilon^2}\delta'(\omega-\omega(p))\) can be
subtracted by counterterms. In section 14 we study Bogoliubov
derivation of kinetic equations and show that the scattering
integral is compensated by some boundary terms which are usually
neglected. Section 15 is a conclusion.
 \section{The algebra of canonical commutative relations}
 The algebra of canonical commutative relations is a \(\star\)
 algebra with a unite generated by generators
 \begin{eqnarray}
 a(f),\;a^+(f),
 \end{eqnarray}
 where \(f\) belongs to the Schwartz space of test functions
 \(S(\mathbb{R}^3)\).
 The generators \(a(f),\;a^+(f)\) satisfies the following relations:\newline
 \(a(f)\) is an antilinear functional on \(f\),\newline
\(a^+(f)\) is a linear functional on \(f\) and\newline
\begin{eqnarray}
{[a(f),a^+(g)]}=\langle f,g\rangle \nonumber\\
{[a(f),a(g)]}={[a^+(f),a^+(g)]}=0.
\end{eqnarray}
Here \(\langle f, g\rangle \) is a standard scalar product in \(L^2(\mathbb{R}^3)\)
\begin{eqnarray}
\langle f,g\rangle=\int f^*(k)g(k)d^3k.
\end{eqnarray}
The involution \(\star\) is defined by the following rule
\begin{eqnarray}
(a(f))^{\star}=a^+(f).
\end{eqnarray}
The "field operators" \(\Psi(x)\;\Psi^+(x)\) are defined by the
following formulas
\begin{eqnarray}
\Psi(x)=\frac{1}{(2\pi)^{\frac{3}{2}}}\int e^{ikx}a(k)d^3k,\nonumber\\
\Psi^+(x)=\frac{1}{(2\pi)^{\frac{3}{2}}}\int e^{-ikx}a^+(k)d^3k.
\end{eqnarray}
Here we have used the following formal notation
\begin{eqnarray}
a^+(f)=\int a(k)f(k)d^3k.
\end{eqnarray}

\textbf{Definition.} Let us consider a Gauss state \(\rho_0\) on
the algebra of canonical commutative relations defined by its
two-point correlator as follows
\begin{eqnarray}
\rho_0(a(k)a(k'))=\rho_0(a^+(k)a^+(k'))=0,\nonumber\\
\rho_0(a^+(k)a(k'))=n(k)\delta(k-k').
\end{eqnarray}
If
\begin{eqnarray}
n(k)=\frac{e^{-\beta\omega(k)}}{1-e^{-\beta\omega(k)}},
\end{eqnarray}
the state \(\rho_0\) is called a Plank state. Here \(\omega(k)=\frac{k^2}{2}-\mu\), \(\mu<0\).
\section{The model}
Our model is described by the following Hamiltonian
\begin{eqnarray}
H=H_0+\lambda V,
\end{eqnarray}
where
\begin{eqnarray}
H_0=\int \omega(k)a^+(k)a(k)d^3k, \nonumber\\
\omega(k)=\frac{k^2}{2}-\mu,\;\mu<0,
\end{eqnarray}
\begin{eqnarray}
V=\frac{1}{2}\int \Psi^+(x)\Psi^+(x')V(x-x')\Psi(x')\Psi(x) d^3x
d^3x',\nonumber\\
\lambda \in \mathbb{R},
\end{eqnarray}
V is an interaction, \(V(x) \in S(\mathbb{R}^3)\). Let us rewrite
the interaction in the Fourier representation
\begin{eqnarray}
V=\frac{1}{2(2\pi)^3}\int d^3k_1 d^3k_2 d^3 k'_1 d^3 k'_2
\tilde{V}(k_1+k_2)\times\nonumber\\
\delta(k_1+k_2-k'_1-k'_2)a^+(k'_1)a^+(k'_2)a(k_1)a(k_2).
\end{eqnarray}
Here by definition
\begin{eqnarray}
\tilde{V}(k)=\int e^{ikx}V(x)d^3x.
\end{eqnarray}
\section{Nonequilibrium diagram technique}
Let us introduce the Green functions for the system
\begin{eqnarray}
\rho(T(\Psi^\pm_H(t_1, x_1),...,\Psi^\pm_H(t_n, x_n))).
\end{eqnarray}
Here symbol \(H\) near \(\Psi^\pm\) means that \(\Psi^\pm_H\) are
Heizenberg operators.

In nonequilibrium diagram technique we admit the following
representation for the Green functions
\begin{eqnarray}
\rho(T(\Psi^\pm_H(t_1,x_1),...,\Psi^\pm_H(t_n,x_n)))=\nonumber\\
\rho(S^{-1}T(\Psi^\pm_0(t_1,x_1),...,\Psi^\pm_0(t_n,x_n)S)).
\end{eqnarray}
Here the symbol \(0\) near \(\Psi^{\pm}\) means that
\(\Psi^{\pm}_0\) are operators in the Dirac representation
(representation of interaction). The \(S\)-matrix has the form
\begin{eqnarray}
S=T\rm exp \mit{(-i\int \limits_{-\infty}^{+\infty}V(t)dt)},\;
\end{eqnarray}
and
\begin{eqnarray}
S^{-1}=\tilde{T}\rm exp \mit{(i\int
\limits_{-\infty}^{+\infty}V(t)dt)}.
\end{eqnarray}
Here \(\tilde{T}\) is a symbol of antichronological ordering.

Let us recall the basics elements of Nonequilibrium diagram
technique. The vertices coming from \(T\)-exponent are marked by
symbol \(-\). The vertices coming from \(\tilde{T}\)-exponent are
marked by symbol \(+\). There exists four tips of propagators
\begin{eqnarray}
G_0^{+-}(t_1-t_2,x_1-x_2)=\rho_0(\Psi(t_1,x_1)\Psi^{+}(t_2,x_2)),\nonumber\\
G_0^{-+}(t_1-t_2,x_1-x_2)=\rho_0(\Psi^+(t_2,x_2)\Psi(t_1,x_1)),\nonumber\\
G_0^{--}(t_1-t_2,x_1-x_2)=\rho_0(T(\Psi(t_1,x_1)\Psi^+(t_2,x_2))),\nonumber\\
G_0^{++}(t_1-t_2,x_1-x_2)=\rho_0(\tilde{T}(\Psi(t_1,x_1)\Psi^+(t_2,x_2))).    \nonumber\\
\end{eqnarray}
Let us write the table of propagators
\begin{eqnarray}
G_0^{+-}(t,x)=\int \frac{d^4k}{(2\pi)^4}(2\pi)\delta(\omega-\omega(k))(1+n(k))e^{-i(\omega t-kx)},\nonumber\\
G_0^{-+}(t,x)=\int \frac{d^4k}{(2\pi)^4}(2\pi)\delta(\omega-\omega(k))n(k)e^{-i(\omega t-kx)},\nonumber\\
G_0^{--}(t,x)=i\int
\frac{d^4k}{(2\pi)^4}\{\frac{1+n(k)}{\omega-\omega(k)+i0}
-\frac{n(k)}{\omega-\omega(k)-i0}\}e^{-i(\omega t-kx)},\nonumber\\
G_0^{++}(t,x)=i\int
\frac{d^4k}{(2\pi)^4}\{\frac{n(k)}{\omega-\omega(k)+i0}
-\frac{1+n(k)}{\omega-\omega(k)-i0}\}e^{-i(\omega t-kx)}.
\end{eqnarray}
\section{Divergences}
A typical example of divergent diagram is pictured on fig1.
\begin{picture}(300,100)
\put(10,80){fig. 1} \put(260,50){\vector(-1,0){45}}
\put(200,50){\oval(30,20)} \put(185,50){\vector(-1,0){15}}
\put(130,50){\vector(-1,0){15}} \put(85,50){\vector(-1,0){45}}
\put(100,50){\oval(30,20)}
\multiput(162,50)(-11,0){3}{\circle*{4}}
\put(260,45){\line(-1,0){6}} \put(218,41){+}
\put(224,39){\line(-1,0){6}} \put(176,41){+}
\put(176,39){\line(1,0){6}} \put(118,41){+}
\put(124,39){\line(-1,0){6}}
 \put(76,41){+}
\put(76,39){\line(1,0){6}} \put(40,45){\line(1,0){6}}
\end{picture}
The ovals represent the sum of one-particle irreducible diagrams. These diagrams are called chain diagrams.
Let us suppose that all divergences of self-energy parts (ovals) are subtracted.
The divergences arises from the fact that singular supports of propagators are coincide.
At first we consider diagrams with one self-energy insertion (one-chain diagram). These diagrams
are pictured on fig. 2.

\begin{picture}(200,100)
\put(10,80){fig. 2}
\put(160,50){\vector(-1,0){45}}
\put(100,50){\oval(30,20)}
\put(85,50){\vector(-1,0){45}}
\put(160,45){\line(-1,0){6}}
\put(118,41){+}
\put(124,39){\line(-1,0){6}}
\put(76,41){+}
\put(76,39){\line(1,0){6}}
\put(40,45){\line(1,0){6}}
\end{picture}
 These diagrams are analogues to one-loop diagrams
in quantum field theory.

There exist two possible types of counterterms. The first one is a counterterms of mass renormalization.
Mass renormalization is equivalent to the following replacement
\begin{eqnarray}
\lambda V\rightarrow\lambda V+M,
\end{eqnarray}
where \(M\) has the form
\begin{eqnarray}
M=\int m(k)a^+(k)a(k)d^3k ,
\end{eqnarray}
\(m(k)\) is a real-valued function of \(k\).

The second type of counterterms are counterterms of asymptotical state. Asymptotical state renormalization
is equal to the following replacement
\begin{eqnarray}
\rho_0(\cdot)\rightarrow \frac{1}{Z}\rho_0(e^{-\int \limits_{-\infty}^{+\infty} h(t)dt}(\cdot)),
\end{eqnarray}
where
\begin{eqnarray}
h=\int h(k)a^+(k)a(k)d^3k,\nonumber
\end{eqnarray}
\(h(k)\) is a real-valued function and
\begin{eqnarray}
Z=\rho_0(e^{-\int \limits_{-\infty}^{+\infty} h(t)dt}).
\end{eqnarray}
We will proof below that the counterterms of asymptotical state
are enough for the renormalization of all one- and two-chain
diagrams.
\section{Proof of the existence of divergences in the theory}

Suppose that there no divergences in Keldysh diagram technique
if \(n(k)\neq \frac{1}{e^{\alpha\frac{k^2}{2}+\beta}+1}\) for any positive \(\alpha,\;\beta\). Therefore
the Green function
\begin{eqnarray}
\rho( S^{-1}(S\Psi^+(t_1,x_1)\Psi(t_2,x_2)))
\end{eqnarray}
is translation invariant. So the density matrix
\begin{eqnarray}
\rho_t(x_1,x_2):=\rho( S^{-1}(S\Psi^+(t,x_1)\Psi(t,x_2)))
\end{eqnarray}
is an integral of motion. Let
\begin{eqnarray}
\rho_t(k)=\int d^3x \rho_t(0,x)e^{ikx}.
\end{eqnarray}
In zero order of perturbation theory \(\rho(k)=n(k)\). But if there no divergences
in Keldysh diagram technique it is possible (see \cite{3}) to derive
the following kinetic equation for \(\rho(k)\)
\begin{eqnarray}
\frac{\partial \rho_t(k)}{\partial t}\nonumber\\
=\int w(p,p_1|p_2,p_3)\{(1+\rho(p))(1+\rho(p_1))\rho(p_2)\rho(p_3)\nonumber\\
-\rho(p)\rho(p_1)(1+\rho(p_2))(1+\rho(p_3))\}.
\end{eqnarray}
The right hand side of this equation is equal to zero only if
\begin{eqnarray}
\rho(k)= \frac{1}{e^{\alpha\frac{k^2}{2}+\beta}+1}
\end{eqnarray}
for some \(\alpha,\beta\) \((\alpha>0,\;\beta>0)\).
But \(n(k)=\rho(k)\) in zero order of perturbation theory, so
\(n(k)\) has a Bose-Einstein form. This contradiction proves our statement.
\section{Regularization} Let us now introduce regularization. Note that
\begin{eqnarray}
\frac{1}{x+i\varepsilon}=\frac{x}{x^2+\varepsilon}-\pi \frac{i}{\pi} \frac{\varepsilon}{x^2+\varepsilon^2}.
\end{eqnarray}
Therefore we use the following regularization
\begin{eqnarray}
\delta(\omega-\omega(k))\rightarrow \frac{1}{\pi} \frac{\varepsilon}{(\omega-\omega(k))^2+\varepsilon^2)}=:
\delta_\varepsilon(\omega-\omega(k)),
\nonumber\\
\mathcal{P}(\frac{1}{\omega-\omega(k)})\rightarrow \frac{{\omega-\omega(k)}}{(\omega-\omega(k))^2+\varepsilon^2)}=:
\mathcal{P}_\varepsilon(\frac{1}{\omega-\omega(k)}).
\end{eqnarray}
\section{Some simple relation on the Green functions}

\textbf{Lemma 1.} The following equalities hold
\begin{eqnarray}
G^{--}(t_1-t_2,x_1-x_2)^\star=G^{++}(t_2-t_1,x_2-x_1),\label{I}\\
G^{+-}(t_1-t_2,x_1-x_2)^\star=G^{+-}(t_2-t_1,x_2-x_1).\label{II}
\end{eqnarray}

\textbf{Proof.} We have
\begin{eqnarray}
G^{--}(t_1-t_2,x_1-x_2)^\star=\rho_0(T(\Psi(t_1,x_1)\Psi^+(t_2,x_2)))^\star\nonumber\\
=\rho_0(\tilde{T}(\Psi^+(t_1,x_1)\Psi(t_2,x_2)))=G^{++}(t_2-t_1,x_2-x_1).
\end{eqnarray}
So the equality \ref{I} is proved. We have
\begin{eqnarray}
G^{+-}(t_1-t_2,x_1-x_2)^\star=\rho_0(\Psi(t_1,x_1)\Psi^+(t_2,x_2))^\star\nonumber\\
=\rho_0(\Psi(t_2,x_2)\Psi^+(t_1,x_1))=G^{+-}(t_2-t_1,x_2-x_1).
\end{eqnarray}
So the equality \ref{II} is proved.

The Lemma is proved.

It is easy to prove the following

\textbf{Lemma 2.} The following equality holds
\begin{eqnarray}
G^{+-}(t,x)=\theta(t)G^{--}(t,x)+
\theta(-t)G^{++}(t,x),\nonumber\\
G^{-+}(t,x)=\theta(t)G^{++}(t,x)+
\theta(-t)G^{--}(t,x)
\end{eqnarray}
Let us introduce the following matrix
\begin{eqnarray}
G=\left \|\begin{array}{cc}
 G^{++}&G^{+-}\\
 G^{-+}&G^{--}\\
\end{array} \right\|.
\end{eqnarray}
Let us introduce the similar matrix for the self-energy operator
\begin{eqnarray}
\Sigma=\left \|\begin{array}{cc}
 \Sigma^{++}&\Sigma^{+-}\\
 \Sigma^{-+}&\Sigma^{--}\\
\end{array} \right\|.
\end{eqnarray}
Dyson equations in Fourier representation has the form
\begin{eqnarray}
G=G_0+G_0\Sigma G.
\end{eqnarray}
We have from these equations that
\begin{eqnarray}
\Sigma=G_0^{-1}-C^{-1},
\end{eqnarray}
or in the matrix form
\begin{eqnarray}
\Sigma=\frac{1}{\rm det \mit G_0}\left \|\begin{array}{cc}
 G_0^{--}&-G_0^{+-}\\
 -G_0^{-+}&G_0^{++}\\
\end{array} \right\|-\frac{1}{\rm det \mit G}\left \|\begin{array}{cc}
 G^{--}&-G^{+-}\\
 -G^{-+}&G^{++}\\
\end{array} \right\|. \label{M}
\end{eqnarray}
It follows from Lemma 1 that
\begin{eqnarray}
G^{++}(\omega, p)=G^{--}(\omega, p)^\star \nonumber\\
G^{+-}(\omega, p)=G^{+-}(\omega, p)^\star, \nonumber\\
G^{-+}(\omega, p)=G^{-+}(\omega, p)^\star.
\end{eqnarray}
Therefore \(\rm det \mit G_0,\; \rm det \mit G\) are real and we
have the following

\textbf{Lemma 3.}
\begin{eqnarray}
\Sigma^{--}(t_1-t_2,x_1-x_2)^\star=\Sigma^{++}(t_2-t_1,x_2-x_1),\label{I'}\\
\Sigma^{+-}(t_1-t_2,x_1-x_2)^\star=\Sigma^{+-}(t_2-t_1,x_2-x_1).\label{II'}
\end{eqnarray}

The following Lemma holds.

\textbf{Lemma 4.}
\begin{eqnarray}
\Sigma^{++}(\omega,p)+\Sigma^{--}(\omega,p)=-\Sigma^{-+}(\omega,p)-\Sigma^{+-}(\omega,p).
\end{eqnarray}

\textbf{Proof.} The statement of lemma follows from the Dyson
equation (\ref{M}) and the following two obvious equalities:
\begin{eqnarray}
G^{++}(\omega,p)+G^{--}(\omega,p)=G^{-+}(\omega,p)+G^{+-}(\omega,p),\nonumber\\
G^{++}_0(\omega,p)+G^{--}_0(\omega,p)=G^{-+}_0(\omega,p)+G^{+-}_0(\omega,p).
\end{eqnarray}

\section{Calculation of the propagators in one-chain approximation}
\textbf{Lemma 5.} The following limit equality holds (in the sense of distributions):
\begin{eqnarray}
\lim_{\varepsilon\rightarrow 0} (\delta_\varepsilon^2(x)-\frac{1}{2\pi\varepsilon}\delta_\varepsilon(x))=0,\nonumber \\
\lim_{\varepsilon\rightarrow 0}
(\frac{1}{\varepsilon}\delta_\varepsilon(x)-\frac{1}{\varepsilon}\delta(x))\rm
=reg \mit,
\nonumber\\
\lim_{\varepsilon\rightarrow 0}
\{\frac{1}{\pi}\frac{1}{x^2+\varepsilon^2}-\frac{1}{\varepsilon}\delta(x)\}=\rm
reg \mit,
\nonumber \\
\lim_{\varepsilon\rightarrow
0}\delta_\varepsilon(x)\mathcal{P}_\varepsilon(\frac{1}{x})\rm reg
=\mit,
\nonumber \\
x \in \mathbb{R}.
\end{eqnarray}
Here reg  means some correct distribution.\nonumber\\
\textbf{Proof.} Let \(f(x)\) be some test function with compact support. We have

\begin{eqnarray}
\int \delta_\varepsilon^2(x)f(x)=\frac{1}{\pi^2} \int \frac{1}{(x^2+\varepsilon^2)^2}f(x)dx\nonumber\\
=\frac{1}{\pi^2}\int \frac{\varepsilon^2}{(x^2+\varepsilon^2)^2}
\{f(0)+xf'(0)+x^2\psi(x)\}dx
\end{eqnarray}
for some smooth bounded function \(\psi(x)\). We have
\begin{eqnarray}
\int \delta_\varepsilon^2(x)f(x)=\frac{1}{\varepsilon \pi^2}
\int \frac{1}{(x^2+1)^2}\{f(0)+\varepsilon x f'(0)+\varepsilon^2x^2\psi(\varepsilon x)\}\nonumber\\
=\frac{1}{\pi^2} \{\frac{1}{\varepsilon} \int
\frac{1}{(x^2+1)^2}dx\}f(0)+O(\varepsilon).
\end{eqnarray}
But
\begin{eqnarray}
\int \frac{1}{(x^2+1)^2}dx=\frac{\pi}{2}.
\end{eqnarray}
So
\begin{eqnarray}
\int \delta_\varepsilon^2(x) f(x)=\frac{1}{2\pi \varepsilon} f(0)+O(\varepsilon).
\end{eqnarray}
So the first equality is proved. One can prove other three equality by the same way.

Therefore we see from the Lemmas 1,2, that we can only consider
the function \(G^{--}(t,x)\). But the function \(G^{--}(t,x)\) can
be represented as a sum of chain diagrams. At first let us
consider the diagrams with one self-energy insertion (one-chain
diagram). We have \(G_\varepsilon^{--}=\sum
\limits_{i,j=\pm}H^{ij}_\varepsilon\), where the diagrams
 for \(H^{ij}_\varepsilon\) are presented at the fig. 2.
  We have the following representation for the divergent parts of these diagrams.
\begin{eqnarray}
(H^{--}_\varepsilon)_{div}(\omega,p)+(H^{++}_{\varepsilon})_{div}(\omega,p)\nonumber\\
=2\pi \Sigma^{--}(\omega,p)n(p)(1+n(p))\frac{1}{\varepsilon}\delta(\omega-\omega(p))\nonumber\\
+2\pi
\Sigma^{++}(\omega,p)n(p)(1+n(p))\frac{1}{\varepsilon}\delta(\omega-\omega(p)).
\end{eqnarray}
We see that the divergent part of these two diagrams is real
(because \(\Sigma^{--}=(\Sigma^{++})^*\)).

It is obvious from previous calculations that the sum of two
possible mass-renormalization diagrams is equal to zero. Let us
consider the singular part of other two diagram presented at fig
3.

\begin{picture}(400,100)
\put(10,80){fig. 3}
\put(340,50){\vector(-1,0){45}}
\put(280,50){\oval(30,20)}
\put(265,50){\vector(-1,0){45}}
\put(340,45){\line(-1,0){6}}

\put(304,45){\line(-1,0){6}}
\put(256,41){+}

\put(220,45){\line(1,0){6}}

\put(160,50){\vector(-1,0){45}}
\put(100,50){\oval(30,20)}
\put(85,50){\vector(-1,0){45}}
\put(160,45){\line(-1,0){6}}
\put(118,41){+}

\put(76,45){\line(1,0){6}}
\put(40,45){\line(1,0){6}}
\end{picture}
We have
\begin{eqnarray}
(H_{\varepsilon}^{-+})_{div}(\omega,p)+(H_{\varepsilon}^{+-})_{div}(\omega,p)\nonumber\\
=\pi(2\pi)(1+2n(p))(1+n(p))\delta_\varepsilon^2(\omega-\omega(p))\Sigma^{-+}(\omega,p)\nonumber\\
+\pi(2\pi)(1+2n(p))n(p)\delta_\varepsilon^2(\omega-\omega(p))\Sigma^{+-}(\omega,p)\nonumber\\
=\pi(1+2n(p))(1+n(p))\frac{1}{\varepsilon}\delta(\omega-\omega(p))\Sigma^{-+}(\omega,p)\nonumber\\
+\pi(1+2n(p))n(p)\frac{1}{\varepsilon}\delta(\omega-\omega(p))\Sigma^{+-}(\omega,p)+O(\varepsilon).
\end{eqnarray}
We see that \((H^{--}_\varepsilon)_{div}(\omega,p)+(H^{++}_\varepsilon)_{div}(\omega,p)\),
\((H^{-+}_\varepsilon)_{div}(\omega,p)\) and \((H^{+-}_\varepsilon)_{div}(\omega,p)\) are real.

We will use the dotted line for lines which connects creation-annihilation operators with operators
coming from the vertex: \(\int h(k) a^+(k)a(k)d^3k\) (see fig. 4).
\begin{picture}(200,100)
\put(10,80){fig. 4}
\put(120,50){\vector(-1,0){5}}
\multiput(160,50)(-10,0){5}{\circle*{2}}
\put(100,50){\oval(30,20)}
\put(45,50){\vector(-1,0){5}}
\multiput(85,50)(-10,0){5}{\circle*{2}}
\put(160,45){\line(-1,0){6}}
\put(118,41){+}
\put(124,39){\line(-1,0){6}}
\put(76,41){+}
\put(76,39){\line(1,0){6}}
\put(40,45){\line(1,0){6}}
\end{picture}
So the divergences in \(G^{--}_\varepsilon=\sum \limits_{i,j=\pm} H^{ij}_\varepsilon\)
can be subtracted  by the following counterterm:
\begin{eqnarray}
h(p)=\Sigma^{++}(\omega,p)+\Sigma^{--}(\omega,p)+\frac{(1+2n(p))}{2n(p)(1+n(p))} \nonumber\\
\times\{(1+n(p))\Sigma^{-+}(\omega,p)+n(p)\Sigma^{+-}(\omega,p)\}.
\end{eqnarray}
We have by using Lemma 4
\begin{eqnarray}
h(p)=\frac{1+2n(p)}{2(1+n(p))n(p)}\times \nonumber\\
\{(1+n(p))\Sigma^{-+}(\omega,p)-n(p)\Sigma^{+-}(\omega,p)\}.
\end{eqnarray}
The left hand side of this equation can be rewritten as follows
(in approximation used in \cite{3})
\begin{eqnarray}
h(p)=\frac{1+2n(p)}{2n(p)(1+n(p))} St(p),
\end{eqnarray}
where \(St(p)\) is a scattering integral.
So \(h(p)\neq 0\) for non-equilibrium matter.
\section{Calculation of propagators in two-chain approximation}
We will calculate the divergent parts of all diagrams \(H_\varepsilon^{ijkl},\;i,j,k,l=\pm\) presented at fig. 5.

\begin{picture}(300,100)
\put(10,80){fig. 5}
\put(235,50){\vector(-1,0){45}}
\put(175,50){\oval(30,20)}
\put(85,50){\vector(-1,0){45}}
\put(100,50){\oval(30,20)}

\put(235,45){\line(-1,0){6}}
\put(193,41){+}
\put(199,39){\line(-1,0){6}}
\put(151,41){+}
\put(151,39){\line(1,0){6}}
\put(160,50){\vector(-1,0){45}}
\put(118,41){+}
\put(124,39){\line(-1,0){6}}
\put(76,41){+}
\put(76,39){\line(1,0){6}}
\put(40,45){\line(1,0){6}}
\end{picture}
We have \(G^{--}_\varepsilon=\sum \limits_{i,j=\pm}H^{ij}+
\sum \limits_{i,j,k,l=\pm}H^{ijkl}_\varepsilon\) in two-chain approximation.

Let us calculate \(H^{----}_\varepsilon(\omega,p)\) (fig. 6). We have

\begin{eqnarray}
H^{----}_\varepsilon(\omega,p)=-(\Sigma^{--})^2\times i \left
\{\frac{1+n(p)}{\omega-\omega(p)+i\varepsilon}-
\frac{n(p)}{\omega-\omega(p)-i\varepsilon} \right \}^3\nonumber\\
=-i(\Sigma^{--})^2 \{
\frac{(1+n(p))^3}{(\omega-\omega(p)+i\varepsilon)^3}
-\frac{(n(p))^3}{(\omega-\omega(p)-i\varepsilon)^3}\nonumber\\
+\frac{3(1+n(p))n(p)^2}{(\omega-\omega(p)+i\varepsilon)(\omega-\omega(p)-i\varepsilon)^2}-
\frac{3(1+n(p))^2n(p)}{(\omega-\omega(p)+i\varepsilon)^2(\omega-\omega(p)-i\varepsilon)}\}.
\end{eqnarray}

But \(\left(\frac{1}{\omega-\omega(p)\pm i\varepsilon}\right)^n\) is a distribution. So we have the
following expression for the singular part of \(H^{----}_\varepsilon\).
\begin{eqnarray}
H^{----}_{\varepsilon}(\omega,p)=-\frac{3in(p)^2(1+n(p))}
{(\omega-\omega(p)+i\varepsilon)(\omega-\omega(p)-i\varepsilon)^2}
\nonumber\\
+\frac{3in(k)(1+n(k))^2}{(\omega-\omega(p)+i\varepsilon)^2(\omega-\omega(p)-i\varepsilon)}+O(1).
\end{eqnarray}
We have
\begin{eqnarray}
\frac{1}{(\omega-\omega(p)+i\varepsilon)^2(\omega-\omega(p)-i\varepsilon)}\nonumber\\
=\frac{1}{(\omega-\omega(p)+i\varepsilon)^2}\{\frac{1}{(\omega-\omega(p)+i\varepsilon)}
+2\pi i \delta_\varepsilon(\omega)\}\nonumber\\
=\frac{2\pi i}{(\omega-\omega(p)+i\varepsilon)^2}\delta_\varepsilon(\omega)+O(1).
\end{eqnarray}
Let \(f(\omega)\) be a test function \(f(\omega) \in S\). We will calculate
\begin{eqnarray}
I_\varepsilon:=\frac{1}{\pi\varepsilon^2} \int \frac{1}{(\Omega+i)^2} \frac{1}{\Omega^2+1} f(\varepsilon\Omega)d\Omega.
\end{eqnarray}
But \(f(\varepsilon\omega)=f(0)+\varepsilon\Omega f'(0)+...\). Substituting this decomposition into
last equation, we find
\begin{eqnarray}
I_\varepsilon=\frac{1}{\pi \varepsilon^2} f(0)\int \frac{1}{(\Omega+i)^2(\Omega^2+1)}d\Omega\nonumber\\
+\frac{1}{\pi\varepsilon} f'(0) \int
\frac{\Omega}{(\Omega+i)^2(\Omega^2+1)}d\Omega+O(1).
\end{eqnarray}
We use the Cauchy theorem for calculation these two integrals. Let us close the integration contour
in the upper half-plane. The integrand has only one pole at the upper half-plane at the point \(\Omega=i\).
Therefore
\begin{eqnarray}
A:=\int \limits_{-\infty}^{+\infty} \frac{1}{(\Omega+i)^2(\Omega^2+1)}d\Omega\nonumber\\
=2\pi i \frac{1}{(\Omega+i)^3}|_{\Omega=i}=2\pi i \times \frac{1}{-8i}=-\frac{\pi}{4}.
\end{eqnarray}
By the same way we find
\begin{eqnarray}
B:=\int \frac{\Omega}{(\Omega+i)(\Omega^2+1)}d\Omega\nonumber\\
=2\pi i \frac{\Omega}{(\Omega+i)^3}|_{\Omega=i}=-\frac{\pi i}{4}.
\end{eqnarray}
So we have
\begin{eqnarray}
I_\varepsilon=-\frac{1}{\varepsilon^2}\frac{1}{4}f(0)-\frac{i}{4\varepsilon}f'(0)+O(1),
\end{eqnarray}
or
\begin{eqnarray}
\left (\frac{1}{\omega-\omega(p)+i\varepsilon} \right )^2\delta_\varepsilon(\omega-\omega(p))\nonumber\\
=-\frac{1}{4\varepsilon^2} f(\omega-\omega(p))+\frac{i}{4\varepsilon}\delta'(\omega-\omega(p))\nonumber\\
+O(1).
\end{eqnarray}
In result
\begin{eqnarray}
\frac{1}{(\omega-\omega(p)+i\varepsilon)^2(\omega-\omega(p)-i\varepsilon)}\nonumber\\
=2\pi i (-\frac{1}{4\varepsilon^2}\delta(\omega-\omega(p))+
\frac{i}{4\varepsilon}\delta'(\omega-\omega(p))+O(1)).\nonumber\\
\frac{1}{(\omega-\omega(p)+i\varepsilon)(\omega-\omega(p)-i\varepsilon)^2}\nonumber\\
=-2\pi i (-\frac{1}{4\varepsilon^2}\delta(\omega-\omega(p))-
\frac{i}{4\varepsilon}\delta'(\omega-\omega(p))+O(1)).
\end{eqnarray}
At last
\begin{eqnarray}
(H^{----}_{\varepsilon})_{div}(\omega,p)=\{\frac{3\pi}{2\varepsilon^2}n(p)(1+n(p))(1+2n(p))\delta(\omega-\omega(p))\nonumber\\
-\frac{6i\pi}{4\varepsilon}n(p)(1+n(p))\delta'(\omega-\omega(p))\}(\Sigma^{--}(\omega,p))^2.
\end{eqnarray}
Let us now calculate the diagrams presented at fig. 6,7.

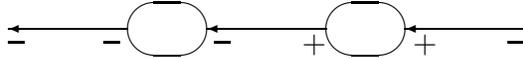
\begin{picture}(300,100)
\put(10,80){fig. 6}
\put(235,50){\vector(-1,0){45}}
\put(175,50){\oval(30,20)}
\put(85,50){\vector(-1,0){45}}
\put(100,50){\oval(30,20)}

\put(235,45){\line(-1,0){6}}

\put(199,45){\line(-1,0){6}}

\put(151,45){\line(1,0){6}}
\put(160,50){\vector(-1,0){45}}
\put(118,41){+}
\put(76,41){+}
\put(40,45){\line(1,0){6}}
\end{picture}

\begin{picture}(300,100)
\put(10,80){fig. 7}
\put(235,50){\vector(-1,0){45}}
\put(175,50){\oval(30,20)}
\put(85,50){\vector(-1,0){45}}
\put(100,50){\oval(30,20)}

\put(235,45){\line(-1,0){6}}
\put(193,41){+}

\put(151,41){+}

\put(160,50){\vector(-1,0){45}}

\put(124,45){\line(-1,0){6}}

\put(76,45){\line(1,0){6}}
\put(40,45){\line(1,0){6}}
\end{picture}
 By the same way as previous we find
\begin{eqnarray}
(H_\varepsilon^{++--})_{div}(\omega,p)=(H_\varepsilon^{--++})_{div}(\omega,p)\nonumber\\
=\frac{2\pi}{4}n(p)(1+n(p))\{\frac{3}{\varepsilon^2}(1+2n(p))\delta(\omega-\omega(p))\nonumber\\
-\frac{i}{\varepsilon}\delta'(\omega-\omega(p))\}\Sigma^{--}(\omega,p)\Sigma^{++}(\omega,p).
\end{eqnarray}

Let us now consider diagrams presented at fig. 8, 9.

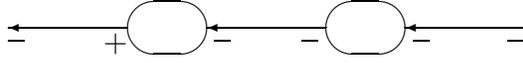
\begin{picture}(300,100)
\put(10,80){fig. 8}
\put(235,50){\vector(-1,0){45}}
\put(175,50){\oval(30,20)}
\put(85,50){\vector(-1,0){45}}
\put(100,50){\oval(30,20)}

\put(235,45){\line(-1,0){6}}

\put(199,45){\line(-1,0){6}}

\put(151,45){\line(1,0){6}}
\put(160,50){\vector(-1,0){45}}
\put(118,41){+}

\put(76,45){\line(1,0){6}}
\put(40,45){\line(1,0){6}}
\end{picture}

\begin{picture}(300,100)
\put(10,80){fig. 9}
\put(235,50){\vector(-1,0){45}}
\put(175,50){\oval(30,20)}
\put(85,50){\vector(-1,0){45}}
\put(100,50){\oval(30,20)}

\put(235,45){\line(-1,0){6}}

\put(199,45){\line(-1,0){6}}

\put(151,45){\line(1,0){6}}
\put(160,50){\vector(-1,0){45}}

\put(124,45){\line(-1,0){6}}
\put(76,41){+}

\put(40,45){\line(1,0){6}}
\end{picture}
 It is easy to see that
\begin{eqnarray}
H^{-+--}_\varepsilon(\omega,p)=\frac{(1+n(p))}{n(p)}\Sigma^{-+}(\omega,p)(\Sigma^{+-}(\omega,p))^{-1}
H^{--+-}_\varepsilon(\omega,p).
\end{eqnarray}
Omitting the calculations we find
\begin{eqnarray}
(H^{-+--}_\varepsilon)_{div}(\omega,p)=(H^{-+--}_\varepsilon)_{div}(\omega,p)\nonumber\\
=2\pi(1+n(p))\Sigma^{-+}(\omega,p)\Sigma^{--}(\omega,p)\nonumber\\
\times\{[\frac{1}{4}((1+n(p))^2+n(p)^2)+n(p)(1+n(p))]
\frac{\delta(\omega-\omega(p))}{\varepsilon^2}\nonumber\\
+\frac{-i}{4\varepsilon}(1+2n(p))\delta'(\omega-\omega(p))\}
\end{eqnarray}
and
\begin{eqnarray}
(H^{--+-}_\varepsilon)_{div}(\omega,p)=(H^{+---}_\varepsilon)_{div}(\omega,p)\nonumber\\
=2\pi n(p)\Sigma^{+-}(\omega,p)\Sigma^{--}(\omega,p) \nonumber\\
\times\{ [\frac{1}{4}((1+n(p))^2+n(p)^2)+n(p)(1+n(p))]
\frac{\delta(\omega-\omega(p))}{\varepsilon^2}\nonumber\\
+\frac{(-i)}{4\varepsilon}(1+2n(p))\delta'(\omega-\omega(p))\}.
\end{eqnarray}
Let us now present analytical expression for other diagrams:
\begin{eqnarray}
(H^{++++}_\varepsilon)_{div}(\omega,p)=\{(1+n(p))n(p)(1+2n(p))\frac{3\pi}{2\varepsilon^2}\delta(\omega-\omega(p))\nonumber\\
+\frac{\pi i}{2
\varepsilon}\delta'(\omega-\omega(p))(1+n(p))n(p)\}(\Sigma^{++}(\omega,p))^2,
\end{eqnarray}
\begin{eqnarray}
(H^{+++-}_\varepsilon)_{div}(\omega,p)=\Sigma^{++}(\omega,p)\Sigma^{+-}
(\omega,p)\frac{\pi}{\varepsilon^2} \nonumber\\
\times(1+3n(p)+3n(p)^2)n(p)\delta(\omega-\omega(p)),
\end{eqnarray}
\begin{eqnarray}
(H^{-+++}_\varepsilon)_{div}(\omega,p)=\Sigma^{-+}(\omega,p)\Sigma^{++}(\omega,p)\nonumber\\
\times(1+n(p))(1+3n(p)+3n(p)^2)
\frac{\pi}{\varepsilon^2}\delta(\omega-\omega(p)),
\end{eqnarray}
\begin{eqnarray}
(H^{+-++}_\varepsilon)_{div}(\omega,p)=
3\pi\Sigma^{+-}(\omega,p)\Sigma^{++}(\omega,p)\delta(\omega-\omega(p))\nonumber\\
\times \frac{(1+n(p))n(p)^2}{\varepsilon^2}, \nonumber\\
\end{eqnarray}
\begin{eqnarray}
(H^{++-+}_\varepsilon)_{div}(\omega,p)=
3\pi\Sigma^{++}(\omega,p)\Sigma^{-+}(\omega,p)\delta(\omega-\omega(p))\nonumber\\
\times\frac{(1+n(p))^2n(p)}{\varepsilon^2}, \nonumber\\
\end{eqnarray}
\begin{eqnarray}
(H^{+--+}_\varepsilon)_{div}(\omega,p)\nonumber\\
=(\frac{2\pi}{4})n(p)(1+n(p))\Sigma^{+-}(\omega,p)\Sigma^{-+}(\omega,p)\nonumber\\
\times n(p)(1+n(p))
\{-\frac{3}{\varepsilon^2}(1+2n(p))\delta(\omega-\omega(p))-
\frac{i}{\varepsilon}\delta'(\omega-\omega(p))\},
\end{eqnarray}
\begin{eqnarray}
(H^{-++-}_\varepsilon)_{div}(\omega,p)=\frac{2\pi}{4\varepsilon^2}\{n(p)^3+2n(p)(1+n(p))^2\nonumber\\
+2n(p)^2(1+n(p))+
(1+n(p))^3\}\nonumber\\
-\frac{2\pi}{4\varepsilon} i\delta'(\omega-\omega(p)) \nonumber\\
\times
\Sigma^{-+}(\omega,p)\Sigma^{+-}(\omega,p)\{1+n(p)+n(p)^2\},
\end{eqnarray}
\begin{eqnarray}
(H^{-+-+}_\varepsilon)_{div}(\omega,p)=-\frac{2\pi}{4}(1+n(p))^2\Sigma^{-+}(\omega,p)\Sigma^{-+}(\omega,p) \nonumber\\
\times
\{-\frac{3}{\varepsilon^2}(1+2n(p))\delta(\omega-\omega(p))+
\frac{i}{\varepsilon}\delta'(\omega-\omega(p))\},
\end{eqnarray}
\begin{eqnarray}
({H^{+-+-}_\varepsilon)_{div}}(\omega,p)=-\frac{2\pi}{4}(n(p))^2\Sigma^{+-}(\omega,p)\Sigma^{+-}(\omega,p) \nonumber\\
\times
\{-\frac{3}{\varepsilon^2}(1+2n(p))\delta(\omega-\omega(p))+
\frac{i}{\varepsilon}\delta'(\omega-\omega(p))\}.
\end{eqnarray}
\section{Counterterm diagrams}
Let us recall that we renormalize the asymptotical state \(\rho\) by the following way
\begin{eqnarray}
\rho()\rightarrow\rho(e^{-\int \limits_{-\infty}^{+\infty}h(t)dt}(\cdot))\frac{1}{Z},
\end{eqnarray}
where
\begin{eqnarray}
Z=\rho(e^{-\int \limits_{-\infty}^{+\infty}h(t)dt}),
\end{eqnarray}
and
\begin{eqnarray}
h(t)=e^{itH_0}he^{-itH_0}.
\end{eqnarray}
So we have to take into account the counterterm diagrams pictured at fig. 10, 11. i.e. \(H^{00\pm\pm}\) and
\(H^{\pm\pm 00}\) respectively.

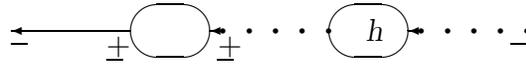
\begin{picture}(300,100)
\put(10,80){fig. 10}
\put(235,50){\vector(-1,0){45}}
\put(175,50){\oval(30,20)}

\put(100,50){\oval(30,20)}

\put(235,45){\line(-1,0){6}}
\put(193,41){+}
\put(199,39){\line(-1,0){6}}
\put(151,41){+}
\put(151,39){\line(1,0){6}}
\put(45,50){\vector(-1,0){5}}
\put(120,50){\vector(-1,0){5}}
\multiput(45,50)(10,0){5}{\circle*{2}}
\multiput(120,50)(10,0){5}{\circle*{2}}
\put(80,39){0}
\put(118,39){0}
\put(40,45){\line(1,0){6}}
\put(99,46){\sl h}
\end{picture}

\begin{picture}(300,100)
\put(10,80){fig. 11}
\put(195,50){\vector(-1,0){5}}
\put(175,50){\oval(30,20)}
\put(85,50){\vector(-1,0){45}}
\put(100,50){\oval(30,20)}
\multiput(120,50)(10,0){5}{\circle*{2}}
\multiput(195,50)(10,0){5}{\circle*{2}}
\put(235,45){\line(-1,0){6}}
\put(174,46){\sl h}
\put(120,50){\vector(-1,0){5}}
\put(118,41){+}
\put(124,39){\line(-1,0){6}}
\put(76,41){+}
\put(76,39){\line(1,0){6}}
\put(40,45){\line(1,0){6}}
\end{picture}
 Let us calculate \(H^{00--}\). We have
\begin{eqnarray}
(H^{00--}_\varepsilon)_{div}=-h(\omega(p),p)(2\pi)^2\delta^2_\varepsilon(\omega-\omega(p))
n(p)(1+n(p)) \nonumber\\
\times
i\Sigma^{--}(\omega,p)\{\frac{1+n(p)}{\omega-\omega(p)+i\varepsilon}-\frac{n(p)}{\omega-\omega(p)-i\varepsilon}\}
\nonumber\\
=-i (2\pi)^2 h(p,\omega(p))\Sigma^{--}(\omega,p)n(p)(1+n(p))\times\nonumber\\
\{[\frac{1}{\varepsilon^2}(\frac{-3i}{8\pi})\delta(\omega-\omega(p))-\frac{1}{\varepsilon}(\frac{1}{8\pi})
\delta'(\omega-\omega(p))](1+n(p))\nonumber\\
-n(p)[\frac{1}{\varepsilon^2}(\frac{3i}{8\pi})\delta(\omega-\omega(p))-\frac{1}{\varepsilon}
\frac{1}{8\pi}\delta'(\omega-\omega(p))]\}.
\end{eqnarray}
In result
\begin{eqnarray}
(H^{00--}_\varepsilon)_{div}(\omega,p)=-h(\omega(p),p)\Sigma^{--}(\omega,p)n(p)(1+n(p))(2\pi)^2\nonumber\\
\times
\{\frac{1}{\varepsilon^2}\frac{3}{8\pi}(1+2n(p))\delta(\omega-\omega(p))-\frac{1}{\varepsilon}
\frac{i}{8\pi}\delta'(\omega-\omega(p))\}.
\end{eqnarray}
Let us present now the analytical expression for \(G^{00+-}\). Omitting some calculation we have:
\begin{eqnarray}
(H^{00+-}_\varepsilon)_{div}(\omega,p)=-h(\omega(p),p)(2\pi)^2\Sigma^{+-}(p,\omega)n(p)(1+n(p))\nonumber\\
\times\{\frac{1}{\varepsilon^2}\delta(\omega-\omega(p))\frac{3}{8\pi}(1+2n(p))-
\frac{i}{\varepsilon}\frac{1}{8\pi}\delta'(\omega-\omega(p))\}.
\end{eqnarray}
Let us present now without calculations all other counterterm
diagrams:
\begin{eqnarray}
(H^{00-+}_\varepsilon)_{div}(\omega,p)=
-h(\omega(p),p)\Sigma^{-+}(\omega,p)\frac{3\pi}{\varepsilon^2}\delta(\omega-\omega(p))\nonumber\\
\times n(p)(1+n(p))^2,
\end{eqnarray}
\begin{eqnarray}
(H^{00++}_\varepsilon)_{div}(\omega,p)=
-\frac{3\pi}{\varepsilon^2}h(\omega(p),p)\Sigma^{++}(\omega,p)\nonumber\\
\times n(p)(1+n(p))^2\delta(\omega-\omega(p)),
\end{eqnarray}
\begin{eqnarray}
(H^{--00}_\varepsilon)_{div}(\omega,p)=-h(\omega(p),p)\Sigma^{--}(\omega,p)n(p)(1+n(p))\nonumber\\
\times(2\pi)^2\{\frac{1}{\varepsilon^2}\delta(\omega-\omega(p))\frac{3}{8\pi}(1+2n(p))-\frac{i}{\varepsilon}
\frac{1}{8\pi}\delta'(\omega-\omega(p))\},
\end{eqnarray}
\begin{eqnarray}
(H^{-+00}_\varepsilon)_{div}(\omega,p)=-h(\omega(p),p)\Sigma^{-+}(\omega,p)n(p)(1+n(p))\nonumber\\
\times(2\pi)^2\{\frac{1}{\varepsilon^2}\delta(\omega-\omega(p))\frac{3}{8\pi}(1+2n(p))-\frac{i}{\varepsilon}
\frac{1}{8\pi}\delta'(\omega-\omega(p))\},
\end{eqnarray}
\begin{eqnarray}
(H^{+\pm00}_\varepsilon)_{div}(\omega,p)=-\frac{3\pi}{\varepsilon^2}n(p)^2(1+n(p))\Sigma^{+\pm}h(\omega(p),p)\nonumber\\
\times\delta(\omega-\omega(p)).
\end{eqnarray}
We have presented all counterterm diagrams.

\section{Annihilation of all strong divergences in the Green functions}
Let \(f(\varepsilon)\) be a function of \(\varepsilon\) of the form
\begin{eqnarray}
f(\varepsilon)=c_{2}\frac{1}{\varepsilon^2}+c_{1}\frac{1}{\varepsilon}+O(1),\;\rm
as\; \mit \varepsilon\rightarrow 0.
\end{eqnarray}
Put by definition
\begin{eqnarray}
(f(\varepsilon))_2=c_{2}\frac{1}{\varepsilon^2}, \nonumber \\
(f(\varepsilon))_1=c_{1}\frac{1}{\varepsilon}.
\end{eqnarray}
We find that \((H^{--++}_\varepsilon)_2\),\((H^{++--}_\varepsilon)_2\),
\((H^{-++-}_\varepsilon)_2\),\((H^{+-+-}_\varepsilon)_2\),\((H^{-+-+}_\varepsilon)_2\),
\((H^{+--+}_\varepsilon)_2\) are real. But the following terms \((H^{----}_\varepsilon)_2\)
and \((H^{++++}_\varepsilon)_2\) are
complex-conjugated to each other.

All not real counterterm diagrams have the form:
\begin{eqnarray}
(H^{00--}_\varepsilon(\omega,p))_2=-(2\pi)^2h(\omega(p),p)n(p)(1+n(p))(1+2n(p))\nonumber\\
\times\frac{3}{8\pi}\frac{1}{\varepsilon^2}\Sigma^{--}(\omega,p)
\delta(\omega-\omega(p)),\nonumber\\
(H^{00++}_\varepsilon(\omega,p))_2=-3\pi h(\omega(p),p)n(p)(1+n(p))^2\nonumber\\
\times\Sigma^{++}(\omega,p)\delta(\omega-\omega(p)),\nonumber\\
(H^{--00}_\varepsilon(\omega,p))_2=-h(\omega(p),p)n(p)(1+n(p))(1+2n(p))\frac{3}{8\pi}(2\pi)^2\nonumber\\
\times\Sigma^{--}(\omega,p)\delta(\omega-\omega(p)),\nonumber\\
(H^{++00}_\varepsilon(\omega,p))_2=-h(\omega(p),p)n(p)^2(1+n(p))\frac{3\pi}{\varepsilon^2}\nonumber\\
\times\Sigma^{++}(\omega,p)\delta(\omega-\omega(p)).
\end{eqnarray}
So it is easy to see that the sum of counterterm diagram is real.

One can see that the sum
\begin{eqnarray}
(H^{++-+}_\varepsilon)_2+(H^{-+++}_\varepsilon)_2+(H^{-+--}_\varepsilon)_2+(H^{---+}_\varepsilon)_2
\end{eqnarray}
is real and
\begin{eqnarray}
(H^{+++-}_\varepsilon)_2+(H^{+-++}_\varepsilon)_2+(H^{+---}_\varepsilon)_2+(H^{--+-}_\varepsilon)_2
\end{eqnarray}
is real too. So all the most strong divergences can be
renormalized by renormalization of the asymptotical state.
\section{Annihilation of all weak divergences in the Green functions}
Now we try to answer the question: if the divergences which are
proportional to \(\delta'(\omega-\omega(p))\) in Green functions
are vanished.

Let us recall what
\begin{eqnarray}
\Sigma^{-+}+\Sigma^{+-}=-\Sigma^{++}-\Sigma^{--}.
\end{eqnarray}
Here and below we omit arguments \((\omega, p)\) of all functions.
We have
\begin{eqnarray}
(H^{----}_\varepsilon)_1=-(\Sigma^{--})^2\frac{6i\pi}{4\varepsilon}n(1+n)\delta'.
\end{eqnarray}
Here and below we will omit an argument of \(\delta\)-function. Corresponding
counterterm is equal
\begin{eqnarray}
(H^{----}_{\varepsilon C})_1=(\Sigma^{--})^2\frac{i \pi}{\varepsilon} n(1+n)\delta'.
\end{eqnarray}
Therefore
\begin{eqnarray}
(H^{----}_{\varepsilon R})_1=-(\Sigma^{--})^2\frac{i\pi}{2\varepsilon}n(1+n)\delta'.
\end{eqnarray}
Here we put by definition
\begin{eqnarray}
H^{\pm\pm\pm\pm}_{\varepsilon R}=H^{\pm\pm\pm\pm}_\varepsilon+H^{\pm\pm\pm\pm}_{\varepsilon C}.
\end{eqnarray}
We have also
\begin{eqnarray}
(H^{++++}_\varepsilon)_1=(\Sigma^{++})^2\frac{i\pi}{2\varepsilon}n(1+n)\delta',
\end{eqnarray}
and corresponding counterterms are equal to zero. In result
\begin{eqnarray}
(H^{----}_{\varepsilon R})_1+(H^{++++}_{\varepsilon R})_1=-\frac{i\pi}{2\varepsilon}\times\nonumber
\{(\Sigma^{--})^2-(\Sigma^{++})^2\}n(1+n)\delta'.
\end{eqnarray}
This quantity is real.

Now let us find counterterms to the diagrams
\begin{eqnarray}
H^{++--}_\varepsilon\; \rm and \mit\; H^{--++}_\varepsilon.
\end{eqnarray}
By using the table of divergent parts of diagrams we find:
\begin{eqnarray}
(H^{++--}_{\varepsilon
C})_1=+\frac{i\pi}{2\varepsilon}\delta'n(1+n)\Sigma^{--}\Sigma^{++}.
\end{eqnarray}
By the same way we find
\begin{eqnarray}
(H^{--++}_{\varepsilon
C})_1=+\frac{i\pi}{2\varepsilon}\delta'n(1+n)\Sigma^{--}\Sigma^{++},
\end{eqnarray}
so
\begin{eqnarray}
(H^{++--}_{\varepsilon R}+H^{--++}_{\varepsilon R})_1=0.
\end{eqnarray}
Now we will calculate the diagram \(H^{-+-+}_\varepsilon\).
\begin{eqnarray}
(H^{-+-+}_\varepsilon)_1=-\frac{2\pi}{4\varepsilon
\varepsilon}(1+n)^2\frac{i}{\varepsilon}(\Sigma^{-+})^2\delta'.
\end{eqnarray}
It is easy to find that its counterterm is equal to
\begin{eqnarray}
(H^{-+-+}_{\varepsilon C})_1=\frac{\pi}{4}\frac{i}{\varepsilon}(1+2n)(1+n)(\Sigma^{-+})^2\delta'.
\end{eqnarray}
Therefore
\begin{eqnarray}
(H^{-+-+}_\varepsilon)_1=-\frac{\pi
i}{4\varepsilon}(1+n)(\Sigma^{-+})^2\delta'.
\end{eqnarray}
Now let us calculate the diagram \(H^{+-+-}\).
\begin{eqnarray}
(H^{+-+-}_\varepsilon)_1=-\frac{2\pi
i}{4\varepsilon}n^2(\Sigma^{+-})^2\delta'
\end{eqnarray}
and
\begin{eqnarray}
(H_{\varepsilon C}^{+-+-})_1=\frac{\pi i}{4\varepsilon}(1+2n)n(\Sigma^{+-})^2\delta'.
\end{eqnarray}
In result:
\begin{eqnarray}
(H_{\varepsilon R}^{+-+-})_1=\frac{\pi i}{4\varepsilon}n(\Sigma^{-+})^2\delta'.
\end{eqnarray}
Now let us consider the diagram:
\begin{eqnarray}
(H^{-+--})_1=\frac{-2\pi
i}{4\varepsilon}(1+n(p))(1+2n(p))\Sigma^{-+}\Sigma^{--}\delta'.
\end{eqnarray}
Let us find counterterms to this diagram:
\begin{eqnarray}
(H^{-+--}_{\varepsilon C})_1=\frac{\pi i}{2 \varepsilon}\Sigma^{-+}\Sigma^{--}(1+n)n\delta'\nonumber\\
+(1+n)(1+2n)\frac{\pi i}{4\varepsilon} \Sigma^{-+}\Sigma^{--}\delta'\nonumber\\
=-\frac{\pi i}{4\varepsilon}(1+n)\Sigma^{-+}\Sigma^{--}\delta'.
\end{eqnarray}
Therefore
\begin{eqnarray}
(H^{-+--}_{\varepsilon R})_1=-\Sigma^{-+}\Sigma^{--}\frac{\pi
i}{4\varepsilon}(1+n)(3+8n)\delta'.
\end{eqnarray}
Let us consider the diagram \(H^{---+}\):
\begin{eqnarray}
(H^{---+}_\varepsilon)_1=-\frac{2\pi
i}{4\varepsilon}(1+n)(1+2n)\Sigma^{-+}\Sigma^{--}\delta'.
\end{eqnarray}
The counterterm corresponding to this diagram is equal to
\begin{eqnarray}
(H^{---+}_{\varepsilon C})_1=\frac{i\pi}{4\varepsilon}(1+n)(1+2n)\Sigma^{-+}\Sigma^{--}\delta'.
\end{eqnarray}

In result
\begin{eqnarray}
(H^{-+--}_{\varepsilon R})_1+(G^{---+}_{\varepsilon R})_1=
\frac{-i\pi}{2\varepsilon}(1+n)^2\Sigma^{-+}\Sigma^{--}\delta'.
\end{eqnarray}
Now let us calculate counterterms to the following diagrams:
\begin{eqnarray}
(H^{+---}_\varepsilon)_1+(H^{--+-}_\varepsilon)_1\nonumber\\
=\frac{\pi(-i)}{\varepsilon}n(1+2n)\delta'.
\end{eqnarray}
\begin{eqnarray}
(H_{\varepsilon C}^{--+-})_1=\pi i\Sigma^{+-}\Sigma^{--}\{\frac{n(1+n)}{2\varepsilon}\nonumber\\
+\frac{n(1+2n)}{4\varepsilon}\}\delta'.
\end{eqnarray}
\begin{eqnarray}
(H_{\varepsilon C}^{+---})_1=\pi i\frac{n(1+2n)}{4\varepsilon}\Sigma^{-+}\Sigma^{--}\delta'.
\end{eqnarray}
Therefore
\begin{eqnarray}
(H_{\varepsilon C}^{--+-})_1+(H_{\varepsilon
C}^{+---})_1=\frac{-i\pi}{2\varepsilon}n^2\Sigma^{-+}\Sigma^{--}\delta',
\end{eqnarray}
and
\begin{eqnarray}
(H^{+---}_{\varepsilon R})_1+(H^{--+-}_{\varepsilon
R})_1=\frac{-i\pi}{2\varepsilon}n^2\Sigma^{+-}\Sigma^{--}\delta'.
\end{eqnarray}
 It is easy to see that:
\begin{eqnarray}
(H^{+-++}_\varepsilon)_1=(H^{-+++}_\varepsilon)_1=
(H^{++-+}_\varepsilon)_1=(H^{+++-}_\varepsilon)_1=0,\nonumber\\
(H_{\varepsilon C}^{+-++})_1=0,\nonumber\\
(H_{\varepsilon C}^{+++-})_1=\frac{i\pi}{\varepsilon}\Sigma^{++}\Sigma^{+-}n(1+n)\delta'.
\end{eqnarray}
So
\begin{eqnarray}
(H_{\varepsilon R}^{+-++}+H_{\varepsilon R}^{+++-})_1=\frac{i\pi}{2\varepsilon}\Sigma^{++}\Sigma^{+-}n(1+n)\delta'.
\end{eqnarray}
It is easy to find that
\begin{eqnarray}
(H^{-+++}_\varepsilon)_1+(H^{++-+}_\varepsilon)_1=0,\nonumber\\
(H_{\varepsilon C}^{-+++})_1=n(1+n)\frac{i\pi}{2\varepsilon}\Sigma^{-+}\Sigma^{++}\delta',\nonumber\\
(H_{\varepsilon C}^{++-+})_1=0.\nonumber\\
\end{eqnarray}
Therefore
\begin{eqnarray}
(H_{\varepsilon R}^{-+++}+H_{\varepsilon}^{++-+})_1=n(1+n)\frac{i\pi}{2\varepsilon}\Sigma^{-+}\Sigma^{++}\delta'.
\end{eqnarray}
Now we must to calculate the following two diagrams:
\begin{eqnarray}
(H^{+--+}_\varepsilon)_1\; \rm and \mit\; (H^{-++-}_\varepsilon)_1.
\end{eqnarray}
Let us start with \((H^{+--+}_\varepsilon)_1\). We have
\begin{eqnarray}
(H^{+--+}_{\varepsilon C})_1=0.
\end{eqnarray}
Therefore
\begin{eqnarray}
(H^{+--+}_{\varepsilon R})_1=-\frac{2i\pi}{4\varepsilon}\Sigma^{-+}\Sigma^{+-}n(p)(1+n(p))\delta'.
\end{eqnarray}
Let us calculate the diagram \(H^{----}_\varepsilon\). We have
\begin{eqnarray}
H^{-++-}_\varepsilon=-\frac{\pi
i}{2\varepsilon}\{1+n+n^2\}\Sigma^{-+}\Sigma^{+-}\delta',
\end{eqnarray}
\begin{eqnarray}
(H^{-++-}_{\varepsilon C})_1=\Sigma^{-+}\Sigma^{+-}\frac{\pi i}{4}((1+2n)(1+2n))\delta'.
\end{eqnarray}
In result
\begin{eqnarray}
(H_{\varepsilon R}^{-++-}+H_{\varepsilon R}^{+--+})=-\frac{\pi i}{4\varepsilon}\Sigma^{+-}\Sigma^{-+}\delta'.
\end{eqnarray}

Now we must summarize all these contribution neglecting by real parts. All real parts can be
subtracted by counterterms of asymptotical state. We have
\begin{eqnarray}
\sum \limits_{i,j,k,l=\pm} (H^{ijkl}_{\varepsilon R})_1\nonumber\\
=\{
-\frac{\pi i}{4}(1+n)(\Sigma^{-+})^2+\frac{\pi i}{4}n(\Sigma^{+-})^2\nonumber\\
-\frac{i\pi}{2}\Sigma^{-+}\Sigma^{--}(1+n)^2-\frac{i\pi}{2}\Sigma^{+-}\Sigma^{--}n^2\nonumber\\
+\frac{\pi
i}{2\varepsilon}\Sigma^{++}\Sigma^{+-}n(1+n)+\frac{i\pi}{2\varepsilon}n(1+n)
\Sigma^{-+}\Sigma^{++}\nonumber\\
-\frac{\pi i}{4\varepsilon}\Sigma^{+-}\Sigma^{-+} \}\delta'.
\end{eqnarray}
Let us unite in the r.h.s. of the last formula 3rd and 6th terms,
and 4th and 5th terms. Neglecting by some real terms we find:
\begin{eqnarray}
\sum \limits_{i,j,k,l=\pm} (H^{i,j,k,l}_{\varepsilon R})_1\nonumber\\
=\{-\frac{\pi i}{4\varepsilon}(\Sigma^{-+})^2+\frac{\pi i}{4\varepsilon}n(\Sigma^{+-})^2\nonumber\\
-\frac{i\pi}{2\varepsilon}\Sigma^{-+}\Sigma^{++}(1+n)
+\frac{\pi i}{2\varepsilon}\Sigma^{++}\Sigma^{+-}n\nonumber\\
-\frac{\pi
i}{4\varepsilon}\Sigma^{+-}\Sigma^{-+}\}\delta'.\label{21}
\end{eqnarray}
It follows from the identity
\begin{eqnarray}
\Sigma^{++}+\Sigma^{--}=-\Sigma^{-+}-\Sigma^{+-}
\end{eqnarray}
that
\begin{eqnarray}
\Sigma^{++}+\Sigma^{-+}=-\Sigma^{--}-\Sigma^{+-},
\end{eqnarray}
and
\begin{eqnarray}
\Sigma^{++}+\Sigma^{+-}=-\Sigma^{--}-\Sigma^{-+}.
\end{eqnarray}
Let us unite in (\ref{21}) the first term with 3rd term and second
term with 4th term. We find
\begin{eqnarray}
\sum \limits_{i,j,k,l=\pm} (H^{ijkl}_{\varepsilon R})_1\nonumber\\
=-\frac{\pi i}{4}\Sigma^{-+}(\Sigma^{++}-\Sigma^{--}-\Sigma^{+-})(1+n)\delta'\nonumber\\
+\frac{\pi i}{4}\Sigma^{+-}(\Sigma^{++}-\Sigma^{--}-\Sigma^{-+})n\delta'\nonumber\\
-\frac{\pi i}{4}\Sigma^{+-}\Sigma^{-+}\delta'. \label{22}
\end{eqnarray}
But \(\Sigma^{+-}\) and \(\Sigma^{-+}\) are real and \((\Sigma^{--})^*=\Sigma^{++}\). Neglecting in
(\ref{22}) by real terms we find
\begin{eqnarray}
\sum \limits_{i,j,k,l=\pm} (H^{ijkl}_{\varepsilon R})_1\nonumber\\
=\{\frac{\pi i}{4}\Sigma^{+-}\Sigma^{-+}(1+n)-\frac{\pi i}{4}\Sigma^{+-}\Sigma^{-+}n\nonumber\\
-\frac{\pi i}{4}\Sigma^{+-}\Sigma^{-+}\}\delta'=0.
\end{eqnarray}
So the image part of divergences of Green function is equal to zero. Therefore Divergences
of two-chain diagram can be subtracted by counterterms of asymptotical state.
 \section{Notes on Bogoliubov derivation of Boltzman equations}
 In this section we study the problem of boundary conditions in Bogoliubov derivation of
 kinetic equations \cite{4}.
Let us consider \(N\) particle in \(\mathbb{R}^3\). Let \({q}_i\)
be a coordinates of particle number \(i\), and \({p}_i\) be a
momenta of particle number \(i\), \(i=1,...,N\). Suppose that
particles interacts by means the pair potential
\(V({q}_i-{q}_j)\). We suppose that \(V\) belongs to the Schwartz
space. Let \(x_i=({p}_i,{q}_i)\) be a point in the phase space
\(\Gamma\). Let \(f(x_1,...,x_n)\) be a distribution function of
\(N\) particle. If we want to point out that \(f(x_1,...,x_N)\)
depends on \(t\) we will write \(f(x_1,...,x_N|t)\). Let
\begin{eqnarray}
f_1(x_1)=\int dx_2...dx_N f(x_1,...,x_n),
\end{eqnarray}
and
\begin{eqnarray}
f_2(x_1,x_2)=\int dx_3,...,dx_N f(x_1,...,x_N)
\end{eqnarray}
be marginal distribution functions. Put by definition
\begin{eqnarray}
\rho_1(x_1)=Nf_1(x_1),\nonumber\\
\rho_2(x_1,x_2)=N^2f_2(x_1,x_2).
\end{eqnarray}
If \(A\) is a function on the phase space \(\Gamma\),
\(\Gamma=\mathbb{R}^{6N}\) and
\begin{eqnarray}
A=\sum \limits_{i=1}^{N}\mathcal{A}(x_i)
\end{eqnarray}
then
\begin{eqnarray}
\langle A\rangle=\int f(x_1,...,x_n) A(x_1,...,x_N)=\nonumber\\
N\int dx_1\mathcal{A}(x_1)f(x_1)=\nonumber\\
=\int dx \mathcal{A}(x)\rho_1(x).
\end{eqnarray}
Now if \(A\) is a function on the phase space
\begin{eqnarray}
A=\sum \limits_{i\neq j}\mathcal{A}(x_i,x_j)
\end{eqnarray}
in the limit of large \(N\) we find
\begin{eqnarray}
\langle A\rangle=\int dx_1 dx_2\rho_2(x_1,x_2)\mathcal{A}(x_1,x_2).
\end{eqnarray}
Let us introduce also three-particle distribution function:
\begin{eqnarray}
f_3(x_1,x_2,x_3)=\int dx_4...dx_N f(x_1,...,x_N).
\end{eqnarray}
Let us derive the equation for \(f(x)\). At first let us write
equation of motion for \(f(x_1,...,x_N)\). We have
\begin{eqnarray}
\frac{\partial}{\partial t}f(x_1,...,x_N\mid t)+\sum
\limits_{i=1}^{n}\frac{{p}_i}{m}{\nabla}_i
f(x_1,...,x_n|t)\nonumber\\
-\sum \limits_{i\neq j} \frac{{\partial} V(q_i-q_j)}{\partial
{q}_i}\frac{\partial f(x_1,...,x_n|t)}{\partial {p}_i}.
\end{eqnarray}
This equation is only an infinitesimal form of the Liouville
theorem. Let us multiple this equation by \(N\) and integrate over
\(dx_2,...,dx_N\). Suppose that \(f(x_1,...,x_N)\) is a function
of rapid decay of momenta. This assumption admit us integrate over
\(p_i\) by parts. We find:
\begin{eqnarray}
\frac{\partial}{\partial
t}\rho_1(x_1|t)+\frac{{p}}{m}{\nabla}\rho_1(x_1,t)\nonumber\\
+\int dx_2 \frac{{p_2}}{m}\frac{\partial}{\partial
{q}_2}\rho_2(x_1,x_2|t)\nonumber\\
=\int dx_2\frac{\partial V(q_1-q_2)}{\partial {q}_1}\frac{\partial
\rho_2(x_1,x_2|t)}{\partial x_1}. \label{0}
\end{eqnarray}
Note that we kept here boundary term. Let us now talk about
derivation of kinetic equation. According to a standard prescription
we put \(\rho_3(x_1,x_2,x_3)=0\) in equation for
\(\rho_2(x_1,x_2)\). We find the following equation for
\(\rho_2\):
\begin{eqnarray}
\frac{d}{dt}\rho_2(x_1(t),x_2(t)|t)=0,\label{1}
\end{eqnarray}
where \((x_1(t),x_2(t))\) is a solution of corresponding two-body
problem.

\textbf{Condition of correlation breaking.} In purpose of
simplicity we consider only translation-invariant matter. Usual
correlation-breaking condition has the form
\begin{eqnarray}
\rho_2(x_1,x_2|0)=h({p}'_1(x_1,x_2))h({p}'_2(x_1,x_2)).
\end{eqnarray}
Here \(h\) is a function on momenta-space of one particle. We
consider only translation-invariant gas, so \(h\) depends only of
momentum.

\({p}_1\) and \({p}_2\) are momenta of particles 1 and 2 at
\(t=-\infty\) if at \(t=0\) their coordinates and momenta was
\(x_1\) and \(x_2\) respectively.

\textbf{Proposition.}
\begin{eqnarray}
\frac{\partial}{\partial t}\rho_2(x_1,x_2)=0.\label{2}
\end{eqnarray}
 Indeed, according to (\ref{1})
 \begin{eqnarray}
 \rho_2(x_1,x_2|t)=\rho_2(x_1^0,x_2^0|o),
 \end{eqnarray}
 where \(x_1^0\) and \(x_2^0\) are phase coordinates of particles
 1 and 2 respectively at a moment \(t=0\). Therefore
 \begin{eqnarray}
\rho_2(x_1,x_2|t)=h({p}'_1(x^0_1,x^0_2))h({p}'_2(x^0_1,x^0_2)).
\end{eqnarray}
But the points \(x_1^0\) and \(x_2^0\) come to the points \(x_1\)
and \(x_2\) after the time t. So \(({p}'_1(x^0_1,x^0_2)
,{p}'_2(x^0_1,x^0_2))=({p}'_1(x_1,x_2) ,{p}'_2(x_1,x_2))\), and
\begin{eqnarray}
\rho_2(x_1,x_2|t)=h({p}'_1(x^0_1,x^0_2))h({p}'_2(x^0_1,x^0_2))=\nonumber\\
h({p}'_1(x_1,x_2))h({p}'_2(x_1,x_2))=\rho_2(x_1,x_2|0).
\end{eqnarray}
In result
\begin{eqnarray}
\rho_2(x_1,x_2|t)=\rho_2(x_1,x_2|0).
\end{eqnarray}
The proposition is proved.

It follows from equations (\ref{1}) and (\ref{2}) that:
\begin{eqnarray}
(\frac{{p}_1}{m}{\nabla}_1+\frac{{p}_2}{m}{\nabla}_2)f(x_1,x_2|t)\nonumber\\
=(\frac{\partial V(q_1-q_2)}{\partial
{q}_1}\frac{\partial}{\partial {p}_1}+\frac{\partial
V(q_1-q_2)}{\partial {q}_2}\frac{\partial}{\partial
{p}_2})f(x_1,x_2|t). \label{3}
\end{eqnarray}
The function \(h(p)\) can be found from the following equation
\begin{eqnarray}
\rho_1(x)=\lim_{N\rightarrow\infty} \frac{1}{N}\int
\rho_2(x_1,x_2) dx_2.
\end{eqnarray}
But in zero order of gas parameter the particles are free and
\begin{eqnarray}
\rho_1(x)=h(x).
\end{eqnarray}
Formula (\ref{3}) is usually used for transformation of r.h.s. of
equation (\ref{0}) to the collision integral. From other hand the
equation
\begin{eqnarray}
\frac{\partial}{\partial t}\rho_2(x_1,x_2)=0
\end{eqnarray}
shows that there no irreversible evolution in the system. From
other point of view we will show that the last term in the left
hand side of (\ref{0}) is equal to the scattering integral.

For simplicity we will show the case \(v_1=0\), \(v=\frac{p}{m}\).
The general case can be reduced to this case by means of Galilei
transformation. So let us consider the integral
\begin{eqnarray}
I=\lim_{R\rightarrow\infty} \int d^3{p}_2 \int
\limits_{V_R}d{q}_2\frac{{p}_2}{m}\frac{\partial}{\partial
{q}_2}\rho(0,0,{p}_2,{q}_2),
\end{eqnarray}
where \(V_R\) is a ball of radius \(R\) with the center at zero.
Let us integrate over \(dq_2\) by using Gauss theorem. We find
\begin{eqnarray}
I=\lim_{R\rightarrow\infty}\int d^3p_2\int \limits_{S_R} dS
\frac{p_2}{m}\cos \psi \rho(0,0,{p}_2,{q}_2).
\end{eqnarray}
Here \(S_R\) is a boundary of \(V_R\) and \(\psi\) is an angle
between to rays: first of them is parallel to \({p}_2\), second
starts from zero and comes throw \(q_2\). We have
\begin{eqnarray}
I=\lim_{R\rightarrow\infty} \int d^3p_2 \int \limits_{S_R}
dS\frac{p_2}{m}\cos\psi\times\nonumber\\
h({p}_1'(0,0))h({p}_2'({p}_2,{q}_2)). \label{4}
\end{eqnarray}
Let us suppose that the particle scatters only then they are not
too far from to each other. Then
\begin{eqnarray}
h({p}_1'(0,0))h({p}_2'({p}_2,{q}_2))=\rho_1({p}_2)\rho_1(0)
\end{eqnarray}
for all \({q}_2 \in S_R\setminus\mathcal{O}\), where
\(\mathcal{O}\) is a small neighborhood of the point
\({q}_0:=\frac{{p}_2}{|{p}_2|}R \in S_R\). Diameter of
\(\mathcal{O}\) is approximately equal to diameter of \(\rm supp
\mit V\). Therefore the integral \(I\) is not equal to zero and
equal to
\begin{eqnarray}
I=\int d^3p_2 \frac{p_2}{m} \int 2\pi b db \nonumber\\
\times\{\rho_1({p}_1'(({p}_2,{q}_2({b})),(0,0))
\rho_1({p}_2'(({p}_2,{q}_2({b})),(0,0))\nonumber\\
-\rho_1({p}_2)\rho_1(0)\}, \label{5}
\end{eqnarray}
where \({b}:={q}_2-{q}_0\). But the right hand side of (\ref{5})
is a usual collision integral.

Therefore if we keep boundary terms in BBGKI-chain we obtain the
kinetic equations without collision integral.
\section{Conclusion}
In this paper we have studied the problem of divergences in
Keldysh diagram technique which arise if the matter is
non-equilibrium. We have considered some wide class of divergent
diagrams and have proposed a method for renormalization of this
divergences. We use this method for renormalization of one- and
two-chain diagrams.

A general thesis that we want to illustrate in this series of papers consists in
follows: to prove that the system tends to thermal equilibrium one should to take
into account its behavior on its boundary. In the last section we have shown that some
boundary terms in BBGKI-chain which are usually neglected in Bogoliubov derivation
of kinetic equation compensate scattering integral in kinetic equation.


\begin{thebibliography}{99999}
\bibitem{1} J.R. Dorfman, E.G. Cohen, Phys. Lett., \textbf{16}, 124 (1965); Journ. Math. Phys.,
\textbf{8}, 282 (1967).
\bibitem{2} K. Kawasaki, I. Oppenheim, Phys. Rev., \textbf{139}, A1763 (1965).
\bibitem{3} E.M. Lifshitz and L.P. Pitaevskii, \textit{Physical Kinetics}, Nauka, Moscow, 1979.
\bibitem{4} N.N. Bogoliubov, \textit{Problems of Dynamical Theory in Statistical Physics}, Gostehizdat, 1946.
\end{thebibliography}
\end{document}